\documentclass[aps,prl,reprint,groupedaddress,showpacs,amsmath,amssymb,twocolumn,floatfix]{revtex4-1}
\usepackage{graphicx}

\begin{document}

\title{Identification of Superfluid Phases of $^3$He in Uniformly Isotropic 98.2$\%$ Aerogel}

\author{J. Pollanen}
\email[]{j-pollanen@northwestern.edu}
\author{J.I.A. Li}
\author{C.A. Collett}
\author{W.J. Gannon}
\author{W.P. Halperin}
\email[]{w-halperin@northwestern.edu}
\affiliation{Northwestern University, Evanston, IL 60208, USA}

\date{\today}

\begin{abstract}
Superfluid $^{3}$He confined to high porosity silica aerogel is the paradigm system for understanding impurity effects in unconventional superconductors.  However, a crucial first step has been elusive; exact identification of the microscopic states of the superfluid in the presence of quenched disorder.  Using a new class of highly uniform aerogel materials, we report pulsed nuclear magnetic resonance experiments that demonstrate definitively that the two observed superfluid states in aerogel are impure versions of the isotropic and axial $p$-wave states.  The theoretically predicted destruction of long-range orbital order (Larkin-Imry Ma effect) in the impure axial state is not observed.           
\end{abstract}

\maketitle

The discovery of effects of quenched disorder on superfluid $^{3}$He using high porosity silica aerogel \cite{Por.95,Spr.95} has created an opportunity for systematic study of the role of impurities on unconventional pairing.  Although the two observed superfluid phases in aerogel have characteristics similar to those of pure $^{3}$He (where the $A$-phase is the axial state and the $B$-phase is the isotropic state), the identification of the states is lacking.  Theory indicates that the presence of elastic quasiparticle scattering reduces strong coupling \cite{Thu.98}, which is known to be responsible for the axial state in pure $^{3}$He.  This should favor the isotropic state in aerogel, consistent with susceptibility and acoustic experiments \cite{Spr.96,Bar.00,Ger.01}. However, a metastable phase is observed at high pressure on cooling, stabilized by magnetic field \cite{Ger.01}, and its identity is more in question.  In this regard we note that without strong coupling, the planar and axial states are degenerate \cite{Vol.90}.  There are predictions that local or global anisotropy in the scattering rates favor various possible anisotropic states, \emph{e.g.} axial \cite{Thu.98}, polar \cite{Aoy.06}, or possibly a family of robust states \cite{Fom.03}.  Furthermore, random local disorder is predicted to lead to an orbitally disordered superfluid glass \cite{Vol.96,Vol.08} or Larkin-Imry-Ma (LIM) state \cite{Lar.70,Imr.75}. To resolve this problem, we have grown a new type of highly homogeneous aerogel and developed methods for its characterization \cite{Pol.08}.  In this Letter we present results on $^{3}$He in a 98.2\% porosity uniformly-isotropic aerogel sample in which we precisely determine the order parameter structure and identify the microscopic states of the superfluid phases to be the impurity suppressed axial and isotropic states.  Additionally, we find no evidence for the existence of the predicted LIM superfluid glass.

Pulsed NMR is a powerful technique for identifying the superfluid states of $^{3}$He, where the frequency shifts of the spectrum are directly related to the amplitude of the order parameter, $\Delta$, and the dependence of the shift on tip angle is a fingerprint of the microscopic state \cite{Vol.90}. However, to date the interpretation of pulsed NMR experiments in aerogel has been complicated by distributions in the frequency shifts owing to spatially non-uniform directions of the angular momentum, called orbital textures, that can be attributed to structural inhomogeneity in aerogel samples on length scales larger than the superfluid coherence length, $\xi_{0}$ \cite{Bar.00,Nak.05,Kun.07,Elb.08,Dmi.10}. 

We performed pulsed NMR measurements on a cylinder of isotropic aerogel, 4.0 mm in diameter and 5.1 mm long.  We grew the 98.2\% porous aerogel sample via the ``one-step'' sol-gel method \cite{Tei.76}.  The sample was characterized with an optical birefringence, cross-polarization technique \cite{Pol.08} with a resolution of $\sim20$ $\mu$m$^{2}$ and found to be uniformly-isotropic at a level $>99.98\%$ \footnote{\emph{I.e.} deviation from uniform isotropy is $<0.02\%$ as measured relative to a known compressed sample.}, indicating a very low level of structural inhomogeneity and anisotropy.  Furthermore, the aerogel was left in the glass tube in which it was grown to avoid strain from differential thermal contractions.  The external magnetic field, $H$, was 31.1, 95.5, or 196 mT (1.008, 3.096, or 6.354 MHz), oriented perpendicular to the aerogel cylinder axis.  Our measurements covered a pressure range from $P=8.1-26.1$ bar.  The sample was cooled via adiabatic nuclear demagnetization of PrNi$_{5}$ to a minimum temperature of 650 $\mu$K.  NMR on $^{195}$Pt was used for thermometry and calibrated relative to the known phase diagram of pure superfluid $^{3}$He from an amount of liquid outside of the aerogel equaling 12.9\% of the total liquid volume.  The geometry of the sample cell used in these experiments is similar to \cite{Spr.95,Spr.96}.  The magnetic susceptibility, $\chi$, was obtained from the numerical integral of the phase corrected absorption spectrum taken from the Fourier transform of the free induction decay, while the NMR frequency shift, $\Delta\omega$, was determined from the power spectrum and measured relative to the bare Larmor frequency, $\omega_{L}$, in the normal state.  

Fig.~1 shows a typical set of susceptibility and frequency shift data taken at $P=26.1$ bar with $H = 196$ mT using a tip angle $\beta=7^{\circ}$ as the sample was warmed with an external heater at a rate of $\sim18$ $\mu$K/hr after demagnetization to zero field. The sharp onset of frequency shifts defines the superfluid transition temperature, $T_{ca}$. The jump in susceptibility indicates the first order transition, $T_{BAa}$, between a $B$-like phase, with a temperature dependent susceptibility, to a field stabilized $A$-like phase having the same susceptibility as the normal state.  Both transitions are extremely narrow ($<3$ $\mu$K wide) consistent with the high uniformity of the aerogel sample as inferred from our optical characterization.
\begin{figure}
\centerline{\includegraphics[height=0.23\textheight]{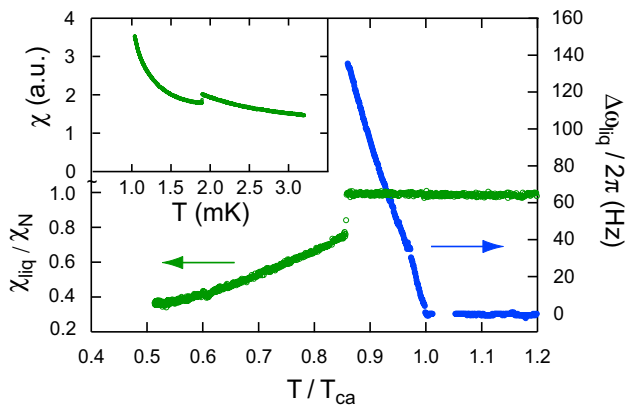}}
\caption{\label{fig1} (Color online) Liquid susceptibility (open green circles) and frequency shift (closed blue circles) versus reduced temperature at $P=26.1$ bar and $H=196$ mT.  The signal from the small amount of liquid outside the aerogel has been subtracted.  The estimated systematic uncertainty in $\chi_{liq}$ is $\sim15\%$ at the lowest temperature.  Inset: Total $^3$He susceptibility versus temperature.}
\end{figure}
The total susceptibility exhibits a well known Curie-Weiss contribution from several layers of solid $^3$He adsorbed to the aerogel surface \cite{Spr.95,Bra.10} and is described by, $\chi=\chi_{sol}+\chi_{liq}=\frac{C}{(T-\theta_{W})}+\chi_{liq}$, where $\chi_{liq}$ is the liquid susceptibility. The background from the solid, $\chi_{sol}$, can be seen in the inset of Fig.~1.  The constant liquid susceptibility in the normal state, $\chi_{N}$, Weiss temperature, $\theta_{W}$, and Curie constant, $C$, were determined over a wide range, $1.0<T<9.1$ mK, giving a temperature dependence of $\chi_{liq}$ (Fig.~1) consistent with the calculation of Sharma and Sauls \cite{Sha.01} for the scattering parameters obtained from the phase diagram of our sample. The $^3$He spins in the liquid and solid phases are in fast exchange \cite{Sch.85,Fre.88,Spr.95,Col.09} and the liquid frequency shift is given by $\Delta\omega_{liq} = \Delta\omega \left(\frac{\chi_{sol}+\chi_{liq}}{\chi_{liq}}\right)$.  The superfluid frequency shifts are due to the quantum coherent dipole torque \cite{Leg.75} and $\Delta\omega_{liq,i}\propto\Omega^{2}_{i}F(\beta,\theta)$, where $\Omega_{i}$ is the longitudinal resonance frequency characteristic of a particular superfluid phase with $i=A,B$ \footnote{Note, $\Omega^{2}_{i} \propto C_{i}\frac{\Delta^{2}_{i}}{\chi_{i}}$ where $\Delta$ is the amplitude of the order parameter, $i=A, B$ labels the superfluid phase, $C_{A}=2/5$, $C_{B}=1$ and the proportionality depends only on normal state properties.}. The factor $F(\beta,\theta)$ depends on tip angle, $\beta$, as well as the angle between the orbital angular momentum and magnetic field, $\theta\equiv(\hat{\ell},\vec{H})$.

The NMR absorption spectra in the normal state, $A$-like and $B$-like phases at $P=26.1$ bar and $H=95.5$ mT are presented in Fig.~2.
\begin{figure}
\centerline{\includegraphics[height=0.2\textheight]{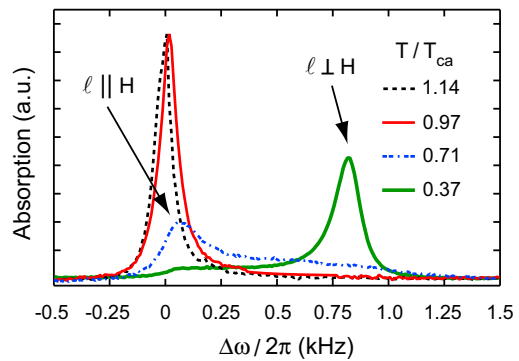}}
\caption{\label{fig2}(Color online) NMR spectra for $^{3}$He in aerogel in the normal state (black dashed), $A$-like (red solid), and $B$-like (blue dashed-dot and green bold solid) phases.  These spectra were obtained in $H=95.5$ mT and at a $P=26.1$ bar where the crossover between $\hat{\ell} \perp$ and $\parallel \vec{H}$ in the $B$-like phase occurs at $T=1.2$ mK and has a width of $\approx100$ $\mu$K.}
\end{figure}
The spectral line width in the $A$-like phase is the same as in the normal state indicating a uniform texture of the superfluid orbital angular momentum, $\hat{\ell}$.  In contrast, for the $B$-like phase near $T_{BAa}$ the spectrum (blue dashed-dot) is comprised of a peak near $\omega_{L}$, with a tail extending to high frequency.  At low temperatures (green bold solid) there is a strong peak at high frequency.  This temperature dependent structure is due to a non-uniform $\hat{\ell}$-texture with the low (high) frequency components corresponding to $\hat{\ell} \parallel \vec{H}$ ($\hat{\ell} \perp \vec{H}$) and we will refer to these as the field- and wall-modes respectively.  We find that the field-mode is dominant near $T_{BAa}$ crossing over to a wall-mode at low temperatures, with the crossover lower at higher field.  The wall-mode has been studied extensively in the isotropic state of pure $^{3}$He \cite{Ish.89,Hak.89,Dmi.99} and arises when $\hat{\ell}$ is oriented perpendicular to $\vec{H}$ by the macroscopic walls of a sample container, as in our case, producing maximal NMR frequency shift.  Similar textures with $\hat{\ell}$ oriented away from $\vec{H}$ have also been observed previously although identification in terms of the isotropic state was not established \cite{Dmi.03,Nak.05}.

The tip angle dependence of the frequency shift, $\Delta\omega_{liq}(\beta)$, is unique to a specific $p$-wave state and can be used to explore the distribution of $\hat{\ell}$.  For the isotropic state, theoretical predictions for the wall and field-modes exist.  Specifically, for the wall-mode the frequency shift as a function of $\beta$ is given by \cite{Dmi.99}
			\begin{eqnarray}
      \Delta\omega_{liq}=\frac{\Omega_{B}^{2}}{2\omega_{L}}(\cos\beta-\frac{1}{5}),\hspace{20pt}\beta < 90^{\circ}\label{1}\\
      \Delta\omega_{liq}=-\frac{\Omega_{B}^{2}}{10\omega_{L}}(1+\cos\beta),\hspace{20pt}\beta > 90^{\circ}\label{2}
      \end{eqnarray}
and for the field-mode we have \cite{Bri.75,Cor.78}
			\begin{eqnarray}
      \Delta\omega_{liq}=0,\hspace{20pt}\beta < 104^{\circ}\label{3}\\
      \Delta\omega_{liq}=-\frac{4}{15}\frac{\Omega_{B}^{2}}{\omega_{L}}(1+4\cos\beta),\hspace{16pt}\beta > 104^{\circ}.\label{4}
      \end{eqnarray}
We have explored the tip angle behavior of the $B$-like phase at $P=26.1$ bar, $H=31.1$, 95.5, and 196 mT, above and below the crossover temperature. These data are presented in Fig.~3 along with fits to Eq.~1-4 as described in the caption.  The excellent agreement between the theory and our measurements confirms that the $B$-like phase in our aerogel is the isotropic state and that we can realize textures with $\hat{\ell}$ $\parallel$ $\vec{H}$ and $\hat{\ell}$ $\perp$ $\vec{H}$.  In the following we will refer to this phase as the $B$-phase.  Additionally, we observe a relatively small shift, independent of $\beta$, for both modes (see Fig.~3).  The origin of this shift is not currently understood but is likely to arise from gap distortion effects similar to those found in the pure $B$-phase \cite{Haa.94,Ran.96}.
\begin{figure}
\centerline{\includegraphics[height=0.21\textheight]{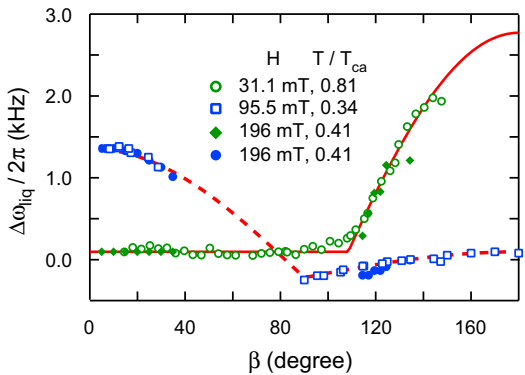}}%
\caption{\label{fig3} (Color online) Frequency shift versus tip angle for the $B$-like phase at $P=26.1$ bar.  The magnitude of the shifts presented have been scaled to a common field of $H=196$ mT.  The solid (dashed) red curves are fits with the theory for the field-mode (wall-mode), Eq.~1-4, using a single fitting parameter, $\Omega_{B}$, obtained from the data in Fig.~4 and with the addition of a small constant shift not contained in the theory and discussed in the text.}
\end{figure}

In the limit of small $\beta$ the frequency shift of the wall-mode reduces to $\Delta\omega_{liq}=\frac{2}{5}\frac{\Omega_{B}^{2}}{\omega_{L}}$ and can be used to determine the temperature dependence of the longitudinal resonance frequency of the $B$-phase.  Our results for $\Omega_{B}^{2}$ at $P=26.1$ bar and $H=196$ mT are presented as the open red circles in Fig.~4.  We also plot the measurements of Hakonen \emph{et al.} \cite{Hak.89} for the longitudinal resonance of the $B$-phase in pure $^3$He (dashed black curve), scaled to account for the impurity suppression of the order parameter and susceptibility in our aerogel, and the agreement with our data (open red circles) is excellent.    
\begin{figure}
\centerline{\includegraphics[height=0.25\textheight]{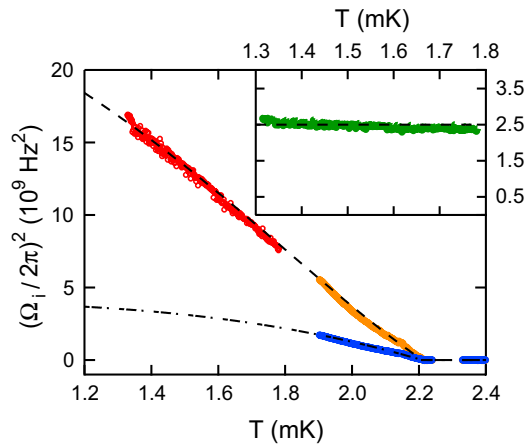}}%
\caption{\label{fig4}(Color online) Temperature dependence of the longitudinal resonance frequency at $P=26.1$ bar and $H=196$ mT in the impurity suppressed isotropic (open red circles) and axial (blue circles, darkest color) states.  The orange data (lightest color) were scaled from the axial state data using $5/2$.  Inset: Leggett ratio versus temperature, Eq.~5.}
\end{figure}

In pure $^{3}$He we have the Leggett relation \cite{Leg.75} uniquely relating the longitudinal resonance frequencies, and hence order parameter symmetries, of the isotropic and axial states in the Ginzburg-Landau regime,
	\begin{equation}
	\frac{5}{2}=\frac{\Omega_{B}^{2}}{\Omega_{A}^{2}}\frac{\chi_{B}}{\chi_{A}}\frac{\Delta_{A}^{2}}{\Delta_{B}^{2}}.\label{5}
	\end{equation}
In Fig.~4 we also present the longitudinal resonance frequency of the high temperature $A$-like phase calculated with the small tip angle expression applicable to the axial state, \emph{i.e}. $\Delta\omega_{liq}=\frac{\Omega_{A}^{2}}{2\omega_{L}}$ (blue circles).  Using Eq.~5 we can scale this data to compare with the isotropic state longitudinal resonance, orange circles (lightest color), in good agreement with the dashed curve.  Conversely, we can use our measurements of the longitudinal resonance and susceptibility of the two phases to calculate the right hand side of Eq.~5 shown in the inset to Fig.~4 which matches well with 5/2.  The analysis in the inset is a comparison of the data for the $B$-phase (open red circles) with the dashed-dot curve in Fig.~4 obtained by extending to low temperature the data of the $A$-like phase (blue closed circles, darkest color) based on the behavior in pure $^{3}$He \cite{Sch.93} that was found earlier to be precisely proportional to that in aerogel \cite{Bau.04}.  We have taken $\Delta_{A}=\Delta_{B}$ because the free energy difference between the two phases is much smaller than the superfluid condensation energy \cite{Bau.04}.  These results demonstrate that the microscopic quantum state of the $A$-like phase is the axial state and we will henceforth call it the $A$-phase.  We have repeated this identification at $P=20.3$ bar with the same conclusion \cite{Li.11}.

The LIM state proposed by Volovik is a consequence of local disorder defined to be on a length scale small compared to $\xi_{0}$, leading to a collapse of the frequency shift and increase in the linewidth \cite{Vol.96,Vol.08}.  Long length scale inhomogeneity can reorient the angular momentum and be difficult to separate from the LIM effect.  The Leggett relation, however, is valid exclusively for the ambient configuration of the isotropic state and the axial state in the dipole-locked texture, where $\hat{\ell} \perp \vec{H}$. Our confirmation of Eq.~5 rules out the existence of the LIM superfluid glass state as proposed by Volovik in our isotropic aerogel.  We have also explored the full extent of supercooling of the axial state in $H=196$ mT down to $T/T_{ca}=0.79$ and find no evidence of orbital disorder in this metastable phase.

Finally in Fig.~5 we present a $P-T$ phase diagram for the impurity suppressed isotropic and axial superfluid states of $^3$He in our isotropic aerogel.  The closed (open) squares are $T_{ca}$ ($T_{BAa}$) and for comparison the phase diagram of pure $^3$He at $H=196$ mT is also shown as the green solid ($T_{c}$) and dashed ($T_{BA}$) curves.  The impurity suppression of the superfluid transition temperature due to aerogel can be understood within the context of the homogeneous isotropic scattering model (HISM) of Thuneberg \emph{et al.} \cite{Thu.98}.  In the HISM this suppression is governed by the quasiparticle mean free path, $\lambda$.  The bold solid red curve in Fig.~5 is a fit to our measured values of $T_{ca}$ using the HISM giving $\lambda=700$ nm, compared to $\sim200$ nm obtained from a cluster aggregation model \cite{Haa.00}.  Furthermore, at $P=26$ bar we have measured $T_{BAa}$ at $H=196$, 95.5 and 31.1 mT, and as $H \rightarrow 0$, we find that $T_{BAa} \rightarrow T_{ca}$, within our experimental accuracy, indicating that the $B$-phase is the stable phase throughout the phase diagram in zero field \cite{Thu.98, Ger.01}.
\begin{figure}
\centerline{\includegraphics[height=0.23\textheight]{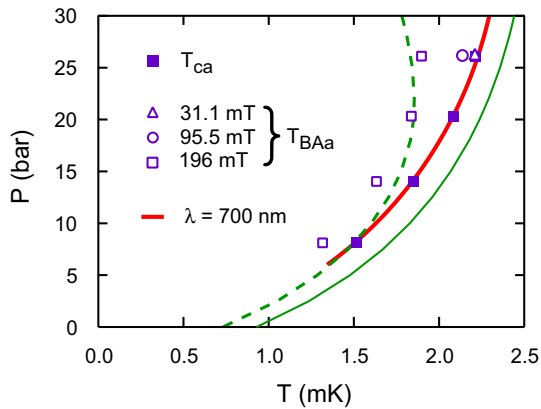}}  
\caption{\label{fig5}(Color online) Superfluid phase diagram for $^3$He in isotropic aerogel.  The closed (open) purple symbols are $T_{ca}$ ($T_{BAa}$ for different fields at $P=26$ bar) and the solid (dashed) green curves correspond to $T_{c}$ ($T_{BA}$) for pure $^3$He at $H=196$ mT.  The bold solid red curve is fit to $T_{ca}(P)$ using the HISM \cite{Thu.98}. The phase diagram can also be fit well with a scattering model that includes particle-particle correlations in the aerogel \cite{Sau.03} with a correlation length $\xi_{a}=0-50$ nm indicating that we are unable to precisely determine $\xi_{a}$.}
\end{figure}

In conclusion, we have used the NMR behavior of superfluid $^3$He in a uniformly-isotropic 98.2\% aerogel to identify the microscopic states of the condensates, finding that they correspond to impurity suppressed versions of the axial and isotropic $p$-wave states of $^3$He.  We also find that the classic $5/2$ Leggett ratio holds for the longitudinal resonance frequencies of the two phases.  Even in the presence of local disorder, inherent to aerogel, the axial state angular momentum remains perpendicular to the magnetic field in the dipole-locked configuration throughout the sample in contrast to theoretical predictions for an orbital glass phase (LIM state).

\begin{acknowledgments}
We are grateful to J.A. Sauls, J.M. Parpia, G.E. Volovik, N. Mulders, A.M. Mounce and Y. Lee for helpful discussion and for support from the National Science Foundation, DMR-1103625.
\end{acknowledgments}

\bibliography{HIArefShortv2}

\end{document}